\documentclass{jfm}

\usepackage{graphicx}
\usepackage{newtxtext}
\usepackage{newtxmath}
\usepackage{natbib}
\usepackage{hyperref}
\usepackage{bm}
\hypersetup{
	colorlinks = true,
	urlcolor   = blue,
	citecolor  = black,
}
\linenumbers

\shorttitle{Flow reversals in liquid metal thermal convection}
\shortauthor{Y. -W. Cao, M. -Z. Ai, L. Chen, J. -C. Yang and M. -J. Ni}

\title{Magnetic modulation of flow reversals in liquid metal thermal convection.}

\author{Yan-Wu Cao\aff{1},
	Ming-Zhu Ai\aff{1}, Long Chen\aff{2}, Juan-Cheng Yang\aff{1} 
	\corresp{\email{yangjc@xjtu.edu.cn}}
	\and Ming-Jiu Ni\aff{1,}\aff{2}
	\corresp{\email{mjni@ucas.ac.cn}}}

\affiliation{\aff{1}State Key Laboratory for Strength and Vibration of Mechanical Structures and School of Aerospace, Xi'an Jiaotong University, Xi'an 710049, China
	\aff{2}School of Engineering Science, University of Chinese Academy of Sciences, Beijing 101408, China}

\begin{document}

	\maketitle
		
	\begin{abstract}
Flow reversals are rarely observed in low-Prandtl-number liquid metal convection due to the fluid's exceptionally high thermal diffusivity. Here, we demonstrate that an external transverse magnetic field can induce such reversals in a quasi-two-dimensional (Q2D) rectangular cell with an aspect ratio ($\it\Gamma$) of $0.2$. Our experimental observations reveal that the system initially exhibits periodic dynamics at the onset of reversals before transitioning to stochastic behavior as the ratio of Rayleigh number ($Ra$) to Hartmann number ($Ha$) increases. This transition is governed by the competition between buoyancy and Lorentz forces, with experimental data showing a linear scaling relationship between $Ra$ and $Ha$ at critical points. We develop a theoretical model that incorporates magnetic field effects in low-Prandtl-number convection to predict the reversal frequencies. These findings provide new insights into how magnetic fields can modulate flow regimes in low-Prandtl-number convection, establishing a controlled framework for investigating reversal dynamics in magnetohydrodynamic systems.

\end{abstract}
	
	
{\bf MSC Codes }  {\it(Optional)} Please enter your MSC Codes here
	
\section{Introduction}
\label{sec:Intro}
Flow reversals are a hallmark of complex nonlinear dynamics and occur widely in natural systems, ranging from convective winds in the atmosphere to the polarity shifts of magnetic fields in the Earth and stars, highlighting their significance in fluid mechanics, geophysics, and astrophysics \citep{gallet_reversals_2012,harrison_behaviour_1966,glatzmaier_role_1999}. These reversal events are typically characterized by the spontaneous, often irregular, reorganization of the large-scale circulation (LSC), and they reflect the intricate interplay between instability, turbulence, and system symmetries. However, despite extensive studies across disciplines, the precise mechanism that triggers reversals and the conditions under which they occur remain open questions \citep{castillo-castellanos_reversal_2016,chen_reduced_2020,suri_predictive_2024}. The inherently intermittent nature of reversals, along with their dependence on global and local flow characteristics, poses major challenges to theoretical modeling and predictive understanding.

The Rayleigh-Bénard (RB) convection system in a quasi-two-dimensional (Q2D) rectangular geometry has emerged as a paradigmatic model for studying reversal dynamics in thermal convection \citep{noto_plume-scale_2024,bodenschatz_recent_2000,lohse_small-scale_2010}. In such configurations, the LSC is confined predominantly within a plane, making its orientation and transformation processes more easily identifiable compared to fully three-dimensional setups. Moreover, the Q2D nature of the system suppresses secondary effects such as torsional or sloshing modes of the LSC \citep{xi_origin_2009,xi_higher-order_2016}, thereby simplifying the dynamics and facilitating detailed flow characterization. This makes the Q2D RB system especially suitable for investigating the statistical and dynamical properties of flow reversals.

Previous results have shown that flow reversals predominantly occur within a limited range of Prandtl numbers \citep{sugiyama_flow_2010,ni_reversals_2015,huang_effects_2016}. For fluids with moderate Prandtl numbers ($Pr \sim $0.7-10), such as water or silicone oil, reversals are frequent and can be directly visualized through techniques like Particle Image Velocimetry (PIV), which has enabled detailed mapping of flow patterns and transitions during reversal events \citep{chen_emergence_2019,xi_azimuthal_2008,zhou_oscillations_2009}. In contrast, for low-Prandtl-number fluids such as liquid metals ($Pr \sim 10^{-2}$), flow reversals are rarely observed. The dominant reason is that thermal diffusion acts much faster than momentum diffusion in such fluids, thereby suppressing the coherent thermal plumes that are essential for destabilizing the LSC and initiating reversals. This inhibition has been confirmed through numerical simulations \citep{sugiyama_flow_2010}. Nonetheless, experimental studies have revealed that under certain conditions, externally imposed forces can revive or alter large-scale dynamics. For example, when a Q2D convection system is subjected to rotation, the resulting Coriolis forces can induce persistent circulations and even spontaneous reversals, suggesting that symmetry-breaking mechanisms driven by external fields play a critical role in shaping LSC behavior \citep{wang_persistent_2023}.

In this work, we focus on a Q2D convection system subjected to a horizontal magnetic field and find that flow reversals can occur in low-$Pr$ fluid. This observation contrasts with the widely accepted understanding that such reversals are strongly suppressed in liquid metals. Our study aims to understand how magnetic fields influence the onset and dynamics of flow reversals in a well-defined laboratory setting. By introducing magnetic modulation as an external control parameter, we explore the possibility of inducing symmetry breaking and triggering reversals, offering new insight into the reversal dynamics under externally modulated conditions.

\section{Experimental system}
\label{sec:Exp_sys}
\begin{figure}
	\centerline{\includegraphics[width=0.8\textwidth]{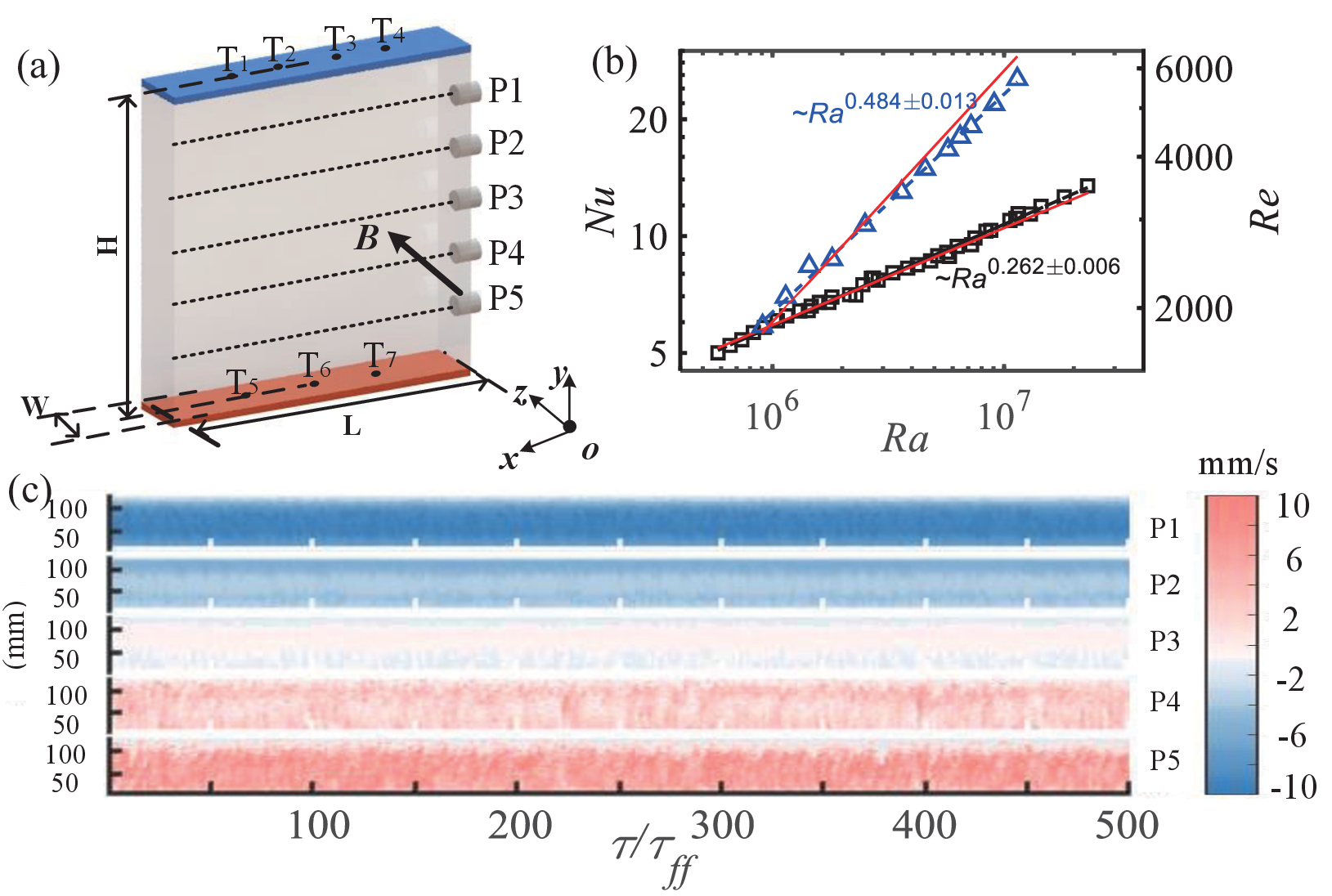}}
	\caption{Probe arrangement and velocity distribution without magnetic field. (a) Arrangement of temperature and ultrasonic probes in the experimental setup. (b) Measured $Nu$ and $Re$: black squares for $Nu$, blue triangles for $Re$, with the dotted line showing the fit and the red line indicating GL theory. (c) Time evolution of velocity at $Ra = 4.54 \times 10^6$, normalized by the free-fall time $\tau_{ff} = \sqrt{H/(\alpha g \Delta T)}$, showing a clear LSC with red and blue denoting opposite flow directions.}
	\label{fig:Fig_1}
\end{figure}

Here, we limit our studies to a Q2D convection system filled with the ternary alloy gallium-indium-tin (GaInSn) with a Prandtl number of $0.029$. Compared to cylindrical or high aspect ratio rectangular convection systems, the flow topologies in Q2D geometry are easily identified. The geometric size of the experimental setup is $12.6 \times 2.52 \times 12.6 $ cm$^3$($L \times W \times H$), as shown in figure \ref{fig:Fig_1}(a), \rm with an aspect ratio of $\it\Gamma=\rm 0.2$. A vertical temperature gradient is applied along the gravitational direction ($y$-axis), while a uniform magnetic field is imposed along the horizontal ($z$-axis), with its intensity varying within $1.2\%$ in the convection region. The dimensionless control parameters are the Rayleigh number $Ra=\alpha g \Delta TH^3/\nu\kappa$ and the Hartmann number $Ha=BH\sqrt{\sigma/(\rho\nu)}$. Here, $\alpha$, $g$, $\Delta T$, $\nu$, and $\kappa$ are the thermal expansion coefficient, gravitational acceleration constant, the temperature difference, kinematic viscosity, and thermal diffusivity of GaInSn at a temperature of $35^{\circ}$C, which is the average of the liquid metal in the convection cell across different cases. The magnetic field strength, height, and conductivity are $B$, $H$, and $\sigma$. The homemade Multichannel Pulsed Ultrasonic Doppler Velocimeter (MPUDV) is adopted to obtain the flow velocity field of liquid metal \citep{pan_development_2024}. As shown in figure \ref{fig:Fig_1}(a), \rm five ultrasonic probes (labeled as P1-P5) are installed on the side wall with a distance of $21$ mm. The MPUDV system can measure the velocity distribution along the $x$ direction on five lines. Meanwhile, multiple thermistors are installed on both the top and bottom plates. The top and bottom plates of the cell are composed of nickel-plated oxygen-free copper, a material that remains non-magnetic and does not interact with the applied magnetic field. Combining the velocity distribution with the temperature difference ($T_4-T_1$ or $T_5-T_7$) measured on the left and right sides of the top and bottom plates, the typical flow characteristics can be identified \citep{ni_reversals_2015,huang_effects_2016}.
	
\section{Results and discussion}
\label{sec:Re&dis}
For the first step, experiments were conducted without magnetic field. Our results demonstrate that the LSC persists throughout the experimental parameter range ($7.20\times10^5< Ra<2.28\times10^7$), in agreement with the findings of \citet{sugiyama_flow_2010}. This is further evidenced in figure \ref{fig:Fig_1}(c), which presents a representative segment of the long-time (120-hour) velocity distribution measurement in GaInSn. Additionally, the scaling relations $Nu \sim Ra^{0.262\pm0.006}$ and $Re \sim Ra^{0.484\pm0.013}$ align well with GL theory predictions \citep{grossmann_scaling_2000,grossmann_prandtl_2002}, as shown in figure \ref{fig:Fig_1}(b).

\begin{figure}
	\centerline{\includegraphics[width=1.0\textwidth]{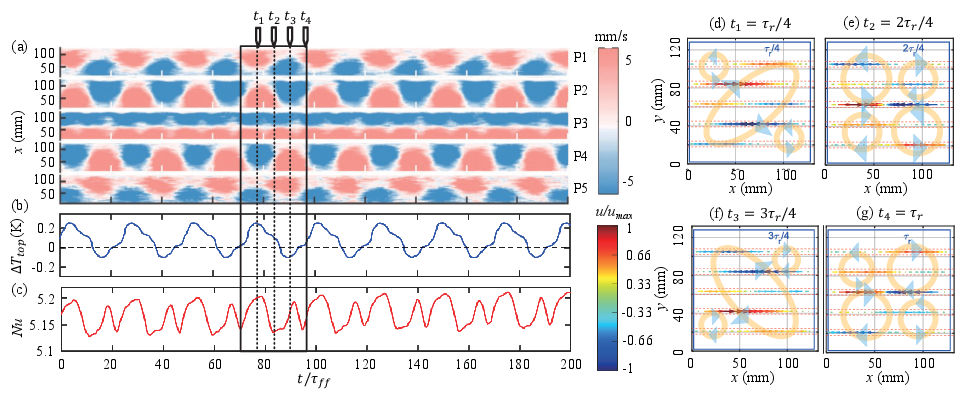}}
	\caption{Typical flow reversal at $Ha=705.39$, $Ra=7.20\times10^6$ from experiment.
		(a) Spatiotemporal velocity distribution: red indicates flow in the positive $x$-direction (away from the probe), blue indicates the opposite.
		(b) Time series of temperature difference $\Delta T_{top} = T_1 - T_4$.
		(c) Instantaneous Nusselt number ($Nu$), synchronized with the velocity field; time is normalized by the free-fall time $\tau_{ff}$. The black box highlights a full reversal cycle, with dashed lines marking $\tau_r/4$, $\tau_r/2$, and $3\tau_r/4$.
		(d)-(g) Cartoons of flow structures at four typical moments within a cycle. Arrows represent velocity vectors derived from MPUDV, with color and size scaled by the instantaneous velocity normalized by $u_{max}$. The $x$ and $y$ axes correspond to directions defined in figure \ref{fig:Fig_1}(a).}
	\label{fig:Fig_2}
\end{figure}

When a horizontal magnetic field was applied, the LSC structure broke down, leading to flow reversals, as shown in figure \ref{fig:Fig_2}(a). At a magnetic field strength of $0.14$ T ($Ha=705.39$), the flow exhibited periodic changes over time. This reversal is further confirmed by temperature measurements, where the temperature difference between adjacent points on the upper wall ($T_4-T_1$) oscillated periodically between positive and negative values. The flow reversal cycle is visualized through typical flow structures in figure \ref{fig:Fig_2}(d)-(g). \rm It is evident that, within the convective cell, the corner rolls located on each of the two diagonals undergo a sequence of processes involving periodic growth, combination, and rupture. These processes subsequently result in flow reversal, a phenomenon analogous to reversals observed in higher-$Pr$ fluids\citep{sugiyama_flow_2010}. However, a key distinction is that the reversal occurs in an extremely low $Pr$ fluid, where the high thermal diffusivity suppresses the development of corner rolls. We perform direct numerical simulations (DNS) to further validate this process, with the numerical methodology detailed in \citet{chen2024effects}. The corresponding flow reversals obtained from experiment and numerical simulation at $Ra=4.54\times10^6$, $Ha=554.23$ can be found in the supplementary videos.

\begin{figure}
	\centerline{\includegraphics[width=0.9\textwidth]{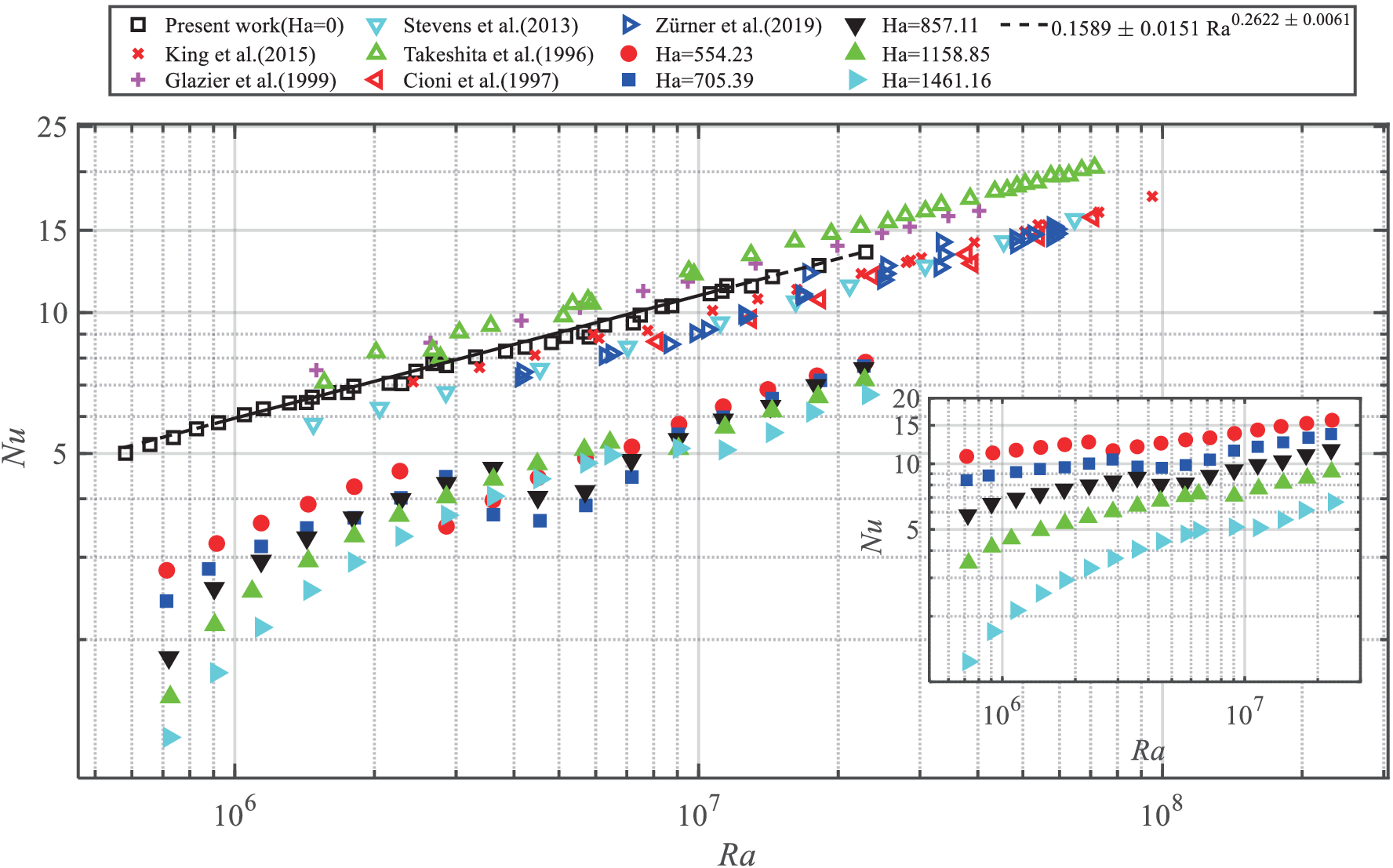}}
	\caption{Heat transfer in liquid metal convection. The scaling of $Nu$ versus $Ra$ is shown. Our data are compared with previous work, as indicated in the legend. The dashed line represents fitted results for $Ha=0$. The solid data points correspond to cases where a magnetic field is applied. For clarity, the $Nu$ values for $Ha =$ 554.23, 705.39, 857.11, 1158.85 and 1461.16 have been vertically shifted by +8, +6, +4, and +2 units, respectively. These artificial vertical shifts are illustrated in the inset plot.}
	\label{fig:Fig_Nu}
\end{figure}

The heat transfer characteristics identified by $Nu$ also exhibited periodic variations with flow structure changes, as shown in figure~\ref{fig:Fig_2}(c). This agrees with our previous work highlighting the strong coupling between flow structure and heat transport in liquid metal thermal convection \citep{chen2023strong}. $Nu$ reaches its maximum when the absolute value of $\Delta T_{top}$ peaks, corresponding to an LSC state with corner rolls. In contrast, heat transfer efficiency is lowest when the LSC structure breaks into multiple rolls. This can be explained by the fact that, in the LSC state, cold and hot plumes effectively transfer heat between the top and bottom plates. However, when the LSC breaks, strong horizontal velocities in the middle of the cell hinder plume movement, significantly reducing heat transfer. The reversal phenomenon also suppresses the mean $Nu$. Figure~\ref{fig:Fig_Nu} compares the measured mean $Nu$ with and without a magnetic field, showing good agreement with prior liquid metal studies when no magnetic field is applied. With a magnetic field, $Nu$ decreases and deviates from the $Ha=0$ case. As shown in the inset of figure~\ref{fig:Fig_Nu}, $Nu$ drops markedly with increasing $Ra$ for each constant magnetic field, reflecting the transition from LSC to reversal. Under the condition of $Ha = 554.23$, the reduction in $Nu$ reaches as much as 23.8\%.

The application of a magnetic field induces flow reversals in the convection system, with experiments showing a transition from periodic to stochastic behavior as $Ra/Ha$ increases. A crucial aspect lies in the magnetic field's control over the transition: it initially enforces periodic reversals, which become stochastic when the buoyancy-to-electromagnetic force ratio surpasses a critical threshold. Following the analytical approach of~\citet{pan2018wake}, we introduce an energy equation incorporating magnetic field effects~\citep{davidson2017introduction}. Since $Pr$ determines the balance between momentum and thermal diffusion, our analysis focuses on how the magnetic field modifies the effective $Pr$. The energy equation with consideration of the magnetic field effects is given by,
\begin{equation}
	\frac{\partial (\bm{u}^2/2)}{\partial t} = -\nabla \cdot(*) + \frac{1}{\rho}(\bm{J}\times \bm{B})\cdot \bm{u} - \nu(\bm{u}\cdot(\nabla \times \bm{\omega}))+g\alpha T \bm{u_y},
	\label{eq:gov1}
\end{equation}

where $*=(p/\rho+u^2/2)\bm{u}$, $\omega$ is vorticity. In comparison to thermal convection in the absence of a magnetic field, an additional energy term $1/\rho (\bm{J}\times \bm{B})\cdot \bm{u}$ is introduced. Using $\delta T$ as the (small) departure of $T$ from the static, linear distribution, we can obtain,
\begin{equation}	
	\frac{\partial (\bm{u}^2/2)}{\partial t} = -\nabla \cdot(*) - \frac{\bm{u}^2}{\sigma \rho} - \nu \bm{\omega}^2 + g\alpha \bm{u_y} \delta T.
	\label{eq:gov2} 	
\end{equation}

For a 2D convection cell, $\bm{J}^2/(\sigma \rho)$ can be expressed as $\sigma B_0^2 \bm{u_x}^2/\rho$ with $\bm{u}$ confined to the $x-y$ plane \citep{davidson2017introduction}, where $B_0$ and $\bm{u_x}$ represent the magnetic field applied along the $z$-axis and the velocity along the $x$-axis respectively. Then the energy equation can be normalized as
\begin{equation}	
	\frac{\partial (\bm{u}^2/2)}{\partial t} = -\nabla \cdot(*) - \nu^{\star} \bm{\omega}^2 + g\alpha \bm{u_y} \delta T.
	\label{eq:gov3}
\end{equation}
Where $\nu^*=\nu(1 + \bm{u_x}^2Ha^2/(\bm{\omega}^2 H^2))$, since $\bm{u_x}^2Ha^2/(\bm{\omega}^2 H^2)$ is always non-negative, the effective viscosity $\nu^{\star}$ is never less than $\nu$, indicating that the effect of introducing an external magnetic field can be analogous to increasing the viscosity of a fluid system.

\begin{figure}
	\centerline{\includegraphics[width=0.9\textwidth]{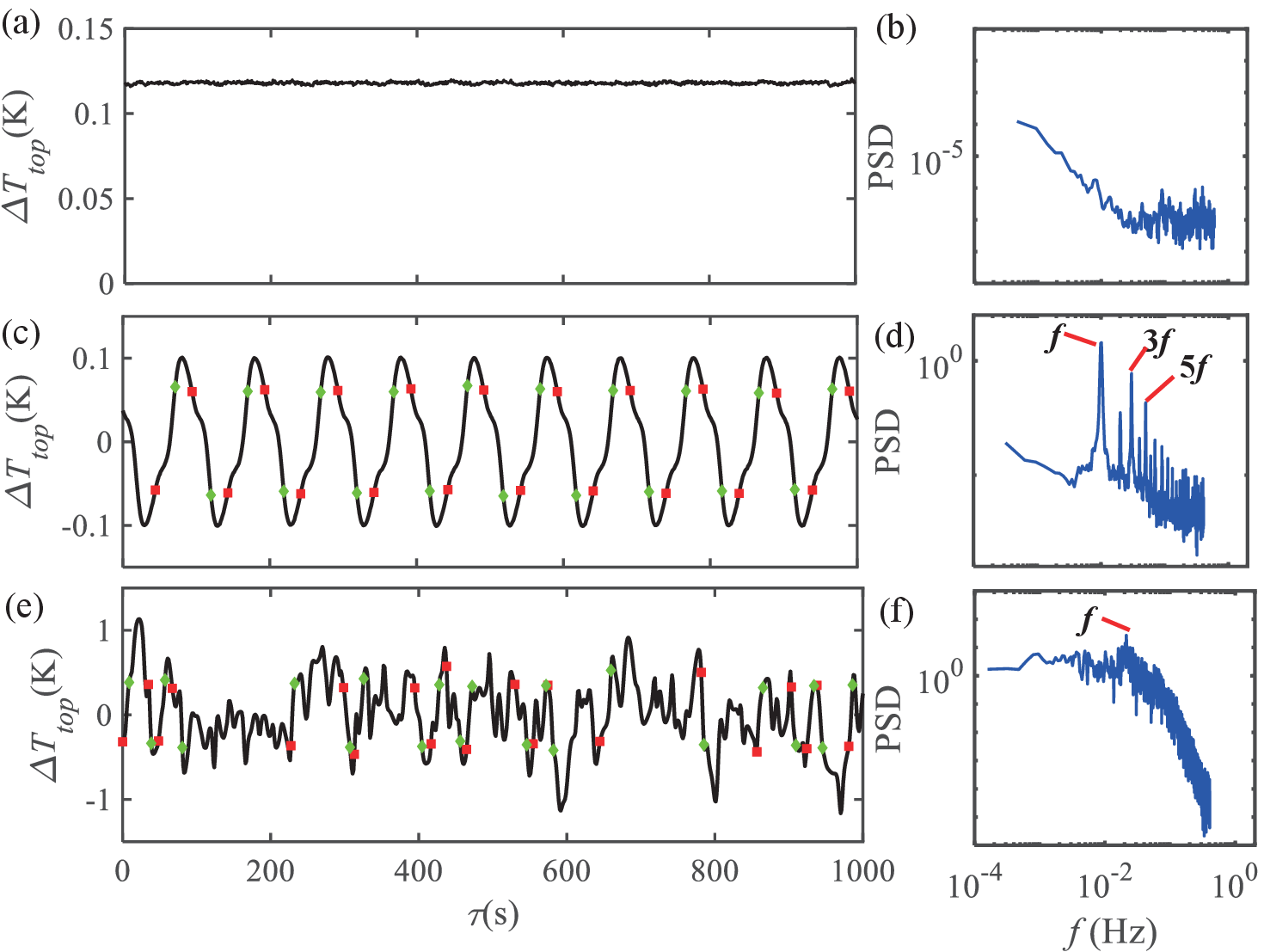}}
	\caption{The reversal characteristics in different regions. (a) and (b) are $\Delta T_{top}$ and its PSD for $Ra=2.86\times10^6$, $Ha=1461.16$ (white region in figure \ref{fig:Fig_5} (a)). (c) and (d) for $Ra=4.54\times10^6$, $Ha=554.23$ (orange region in figure \ref{fig:Fig_5} (a)). (e) and (f) for $Ra=2.28\times10^7$, $Ha=1158.85$ (blue region in figure \ref{fig:Fig_5} (a)). Red squares and green diamonds mark the start and end of reversals.}
	\label{fig:Fig_4}
\end{figure}

The application of an external magnetic field increases the effective viscosity ($\nu^*$), leading to an increase in the effective Prandtl number ($Pr^*$). This modification strengthens viscous forces relative to thermal diffusivity, shifting the system towards a regime where flow reversals become more likely. In low $Pr$ fluids, such as liquid metals, the magnetically induced rise in viscosity perturbs the force balance, promoting the occurrence of flow reversals. To validate the enhancement of $Pr^*$ under magnetic field effects, we use the data from DNS. We evaluate the mean horizontal velocity over the entire $x-y$ plane during 200 seconds after the flow reaches a steady state. First, a spatial average of the horizontal velocity is performed. Since the velocity values vary in sign across different spatial locations, we take the absolute value of the velocity before computing the spatial average. Then, a time average of the spatially averaged horizontal velocity is taken. When the $Ha$ is 550 and the $Ra$ is $4.50\times10^6$, the average horizontal velocity $u_x$ in the 2D plane is found to be $0.0020\ m/s$, and the average vorticity $\omega$ is $0.4835 \ rad/s$. The calculated effective $Pr$ is 9.1215, which falls within the $Ra-Pr$ phase diagram proposed by \citet{sugiyama_flow_2010}, where flow reversals can be observed.

\begin{figure}
	\centerline{\includegraphics[width=1.0\textwidth]{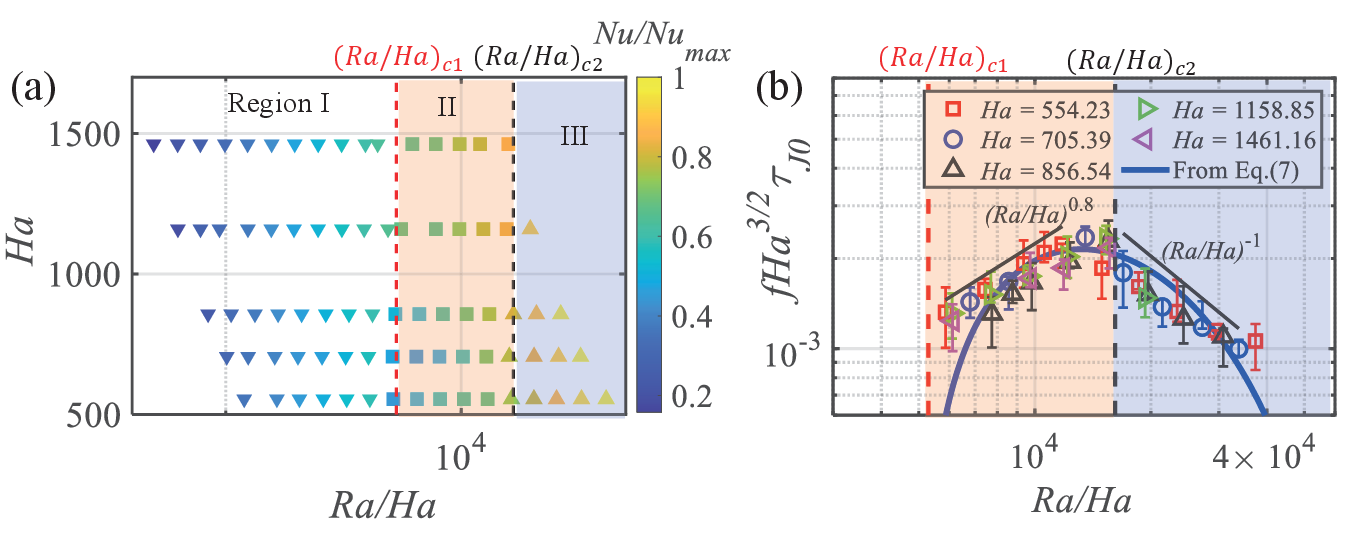}}
	\caption{The statistical characteristics of flow reversal. (a) Phase diagram with three regimes divided by $(Ra/Ha)_{c1} \simeq 5250$ and $(Ra/Ha)_{c2} \simeq 16750$: Region I (stable LSC, inverted triangles), Region II (periodic reversals, squares), Region III (stochastic reversals, upright triangles). The marker color shows normalized $Nu$. Dimensionless reversal frequency with error bars. The error bar originates from the statistics of the reversals. Regions and vertical lines match figure \ref{fig:Fig_5} (a).}
	\label{fig:Fig_5}
\end{figure}

Having established preliminary insights into the physical mechanisms underlying magnetic-field-induced flow reversals, we now turn to the identification and analysis of reversal events. Reversal events are identified by the temperature difference $\Delta T_{top}=T_1-T_4$, following \citet{sugiyama_flow_2010, chen_emergence_2019, huang_effects_2016}. The criteria we used to identify the reversal event from the time trace of $\Delta T_{top}$ are adopted from \citet{huang_effects_2016}. Figure \ref{fig:Fig_4}(a) \rm depicts the state in Region I of figure \ref{fig:Fig_5}(a), \rm where $\Delta T_{top}$ stays nearly constant and positive, indicating a stable LSC, consistent with MPUDV results. The corresponding PSD shows no dominant frequency, similar to the convection regime in \citet{ren_flow_2022}. Figure \ref{fig:Fig_4}(c) \rm shows the condition in Region II, where the flow transitions from a stable LSC to periodic reversals. The PSD in figure \ref{fig:Fig_5}(a) \rm reveals the dominant and harmonic frequencies. Reversals occur when the buoyancy-to-electromagnetic force ratio exceeds a critical value, marked by the red vertical line at $Ra/Ha_{(c1)}$ in figure \ref{fig:Fig_5}(a). \rm Furthermore, $Nu$ drops markedly during LSC reversal transitions, reflected by a clear blue color shift in the associated data points. Figure \ref{fig:Fig_4}(e) \rm shows a typical condition in Region III. The reversals become stochastic, and the corresponding PSD becomes noisy in the full range of frequencies, similar to the PSD of temperature in the turbulent regime of \citet{ren_flow_2022}. As $Ra/Ha$ increases, the flow transitions from stable LSC to periodic reversals and then stochastic reversals.

For the periodic and stochastic reversals observed in the experiment, the transition between these regimes occurs at a critical threshold. According to Burr and M\"{u}ller's linear stability analysis of the critical $Ra_c$ for the oscillation of liquid metal thermal convection under a magnetic field, $Ra_c=4\pi^2/\tau$, where $1/\tau=Q^{1/2}+Qc_H/(b+c_H)$ and $Q$ is the Chandrasekhar number $(Q=Ha^2)$. For electrically insulating Hartman walls, the wall conductance ratio $c_H=\sigma_WS/(\sigma b)\approx 0$, where $\sigma_WS$ is the electrical conductivity of the wall material, $S$ is the thickness of the wall and $b$ is the half-width of the flow region in the direction of the magnetic field, it satisfied $Ra_c\sim Ha$ \citep{burr_rayleighbenard_2001,burr_rayleighbenard_2002}. The critical $Ra$ proposed by \citet{burr_rayleighbenard_2002} represents the threshold at which convection begins to exhibit periodic oscillations. In our experiments, these oscillations manifest as flow reversals. The critical $Ra$ at which reversals first occur shows a linear relationship with the $Ha$, consistent with the scaling predicted by \citet{burr_rayleighbenard_2002}. Furthermore, our experiments reveal that the critical $Ra$, at which the periodic reversal transfers to stochastic reversal, also exhibits a linear relationship with $Ha$. In different reversal regimes, the reversal frequency exhibits distinct variations with $Ra/Ha$. In the following sections, these frequency variations are examined through both theoretical and experimental approaches.

We extended the model based on the work of~\citet{ni_reversals_2015}, which successfully predicted the reversal frequency in water. Considering the work of~\citet{ni_reversals_2015} on the influence of varying aspect ratios on flow reversal behavior, they incorporated convective and diffusive terms into the two-dimensional Brown-Ahlers (B-A) model \citep{brown_large-scale_2007,brown_model_2008}. The convective term is approximated as $-\bm{u} \cdot \nabla T \sim V_c\nu Re\delta/H^2$, where $V_c$ is a prefactor that represents the geometrical coefficient from volume averaging and $\delta$ represents the absolute value of $\Delta T_{top}$. The thermal diffusion term is given by $\kappa\nabla^2 T$. In the process of approximating the thermal diffusion term, they assumed that $N_{pl}=Nu\Gamma^\alpha$, with $\alpha>0$ for simplicity, where $N_{pl}$ is the total number of thermal plumes. In our study, we focus on the influence of different $Ha$ on flow reversal behavior. Analogously, the applied magnetic field inhibits three-dimensional flow structures, promoting the development of Q2D flow characteristics. In simple terms, increasing magnetic field strength leads to a more pronounced Q2D flow organization \citep{burr_rayleighbenard_2002,yang_transition_2021,vogt_free-fall_2021,chen2024effects}. Accordingly, we assume a scaling relation of $N_{pl}=NuHa^\beta$, with $\beta<0$. Following the method of \citet{ni_reversals_2015}, we can obtain
\begin{equation}
	\kappa\nabla^2 T \approx C\kappa\Delta T Nu^2Ra^{0.23}Ha^\beta/H^2,
\end{equation}
where $C$ is also a prefactor that represents the geometrical coefficients from volume averaging. Finally, we formulate an extended Ni model that accounts for both convective and thermal diffusion terms, given by,
\begin{equation}
	\dot{\delta} =-A\delta+B+f_\delta(t),
\end{equation}
where
\begin{gather}
	A=\frac{V_c\nu Re(1-10Re^{-0.5})}{H^2}, B=\frac{C\kappa\Delta TNu^2Ra^{0.23}Ha^b}{H^2} 
	\nonumber	
\end{gather}

Given the probability distribution $p(\delta)$ of $\delta$, the reversal frequency can be derived using the backward Fokker-Planck equation as:
\begin{equation}	
	f=\frac{1}{C_p}exp(\frac{B^2}{AD_{\delta}}),
\end{equation}
where $C_p=\sqrt{2\pi D_\delta/A}/B=2\sqrt{2\pi}\sigma^3/(dD_\delta)$, $D_{\delta}$ describes diffusion and satisfies $D_{\delta}(H^2/(\Delta T^2 \nu))\sim Ra^{0.43}$. Taking into account experimental observations and the scale of $Nu\sim Ra^{0.262}$ and $Re\sim Ra^{0.484}$, the final dimensionless form of the reversal frequency is, 
\begin{equation}	
	fHa^{3/2}\tau_{J0} = m(\frac{Ra}{Ha})^{3/2}exp(-n(\frac{Ra}{Ha})^{0.53}),
\end{equation}
where 
\begin{gather}
	m=\frac{L^2\nu^{-1}Ha^{1/2}}{C_p}, n=\frac{C^2\kappa^2Ha^{0.53+b}}{V_c\nu^2((108/V_c)Re^{-0.5}-1)} 
	\nonumber
\end{gather}

In figure \ref{fig:Fig_5}(b), the normalized reversal frequencies are plotted as $Ra/Ha$. Here, reversal frequencies are non-dimensionalized by the combination of $Ha$ and the Joule dissipation time $\tau_{J0}=\rho/\sigma B^2$. It can be seen that the experimentally obtained reversal frequency is consistent with the theoretical prediction. When the magnetic field is relatively strong, studies by~\citet{kraichnan_inertial-range_1965} indicate that the dissipation mechanism of the system is enhanced, suppressing the turbulence energy spectrum in the high-frequency range. As a result, stochastic effects are weak, and the flow becomes more orderly and stable, with reversals predominantly exhibiting near-periodic behavior. When the magnetic field is weaker, stochastic effects become more pronounced, and reversals exhibit randomness. The critical point between the observed periodic and stochastic reversals occurs at $(Ra/Ha)_{c2}\simeq16750$ from the experiment. The present model successfully predicts this experimental result. The electromagnetic force weakens the stochastic effects in the flow, causing the flow reversals to transition from randomness to periodicity. When the magnetic field is sufficiently strong, the flow sustains a stable LSC. The transition of flow states depends on the interplay between buoyancy and electromagnetic forces. In the experiments, the two transition points exhibit a linear relationship between the critical $Ra$ and $Ha$.
\section*{Conclusion}
In conclusion, we have experimentally demonstrated the existence of a stable LSC in a Q2D convection cell filled with a low Prandtl number liquid metal. Without an external magnetic field, the LSC remains steady and persistent. However, with the influence of an external magnetic field, the flow reversals appear in a well-defined parameter space, where the competition between buoyancy and electromagnetic forces dictates a transition between periodic and stochastic regimes. 

We identify two critical points that characterize the onset of reversals and the transition from periodic to stochastic behavior. At both thresholds, $Ra$, and $Ha$ exhibit a linear relationship, consistent with theoretical expectations. The flow initially exhibits periodic reversals, which become stochastic as $Ra/Ha$ increases. In the periodic regime, the reversal frequency rises with $Ra/Ha$ but decreases after the transition to stochastic reversals. A theoretical model has been developed to describe reversals in low $Pr$ fluid under the influence of a magnetic field. This model predicts reversal frequencies, both periodic and stochastic, that align closely with experimental measurements. The electromagnetic force acts to stabilize and organize the flow, whereas the buoyancy force amplifies its stochasticity. The present findings have important implications for controlling convective behavior in low $Pr$ fluids, particularly concerning flow reversal.
\hspace*{\fill}	

\section*{Acknowledgements} 
The authors gratefully acknowledge the support from NSFC ($\#$52222607, $\#$52176086, $\#$51927812), National Key Research and Development Program of China ($\#$2022YFE03130000), Young Talent Support Plan of Xi'an Jiaotong University.
	
\bibliographystyle{jfm}
\bibliography{jfm}

\end{document}